\documentstyle[12pt]{article}
\advance \topmargin by -\headheight
\advance \topmargin by -\headsep

\evensidemargin \oddsidemargin
\begin{document}
\topmargin -.6in
\def\br{\begin{eqnarray}}
\def\er{\end{eqnarray}}
\def\be{\begin{equation}}
\def\ee{\end{equation}}
\def\nn{\nonumber}
\def\({\left(}
\def\){\right)}
\def\a{\alpha}
\def\b{\beta}
\def\d{\delta}
\def\D{\Delta}
\def\eps{\epsilon}
\def\g{\gamma}
\def\G{\Gamma}
\def\h{ {1\over 2}  }
\def\hp{ {+{1\over 2}}  }
\def\hm{ {-{1\over 2}}  }
\def\k{\kappa}
\def\l{\lambda}
\def\L{\Lambda}
\def\m{\mu}
\def\n{\nu}
\def\o{\over}
\def\O{\Omega}
\def\p{\phi}
\def\rh{\rho}
\def\s{\sigma}
\def\t{\tau}
\def\th{\theta}
\def\ii {\'\i  }
\begin{titlepage}
To appear in Modern Physics Letters A\\
hep-th/9507132
\vskip .6in

\begin{center}
{\large {\bf Supersymmetry, Variational Method and Hulth\'en Potential
}}
\end{center}

\normalsize
\vskip .4in

\begin{center}
{Elso Drigo Filho\footnotemark
\footnotetext{Work supported in part by FAPESP, CNPq and FUNDUNESP}} \\
\par \vskip .1in \noindent
Instituto de Bioci\^encias, Letras e Ci\^encias Exatas-UNESP\\
Departamento de F\ii sica \\
Rua Cristov\~ao Colombo, 2265\\
15055 S\~ao Jos\'e do Rio Preto, SP, Brazil\\
\par \vskip .3in

\end{center}

\begin{center}
{Regina Maria  Ricotta}
\par \vskip .1in \noindent
Faculdade de Tecnologia de S\~ao Paulo, CEETPS-UNESP\\
Pra\c ca Fernando Prestes, 30 \\
01121-060 S\~ao Paulo, SP, Brazil \\
\par \vskip .3in

\end{center}

\begin{center}
{\large {\bf ABSTRACT}}\\
\end{center}
\par \vskip .3in \noindent

The formalism of Supersymmetric Quantum Mechanics provides us the
eigenfunctions to be used in the variational method to obtain the eigenvalues
for the Hulth\'en Potential.
\vskip 8cm
PACS No. 03.65.
\end{titlepage}

\section{Introduction}

The Hulth\'en Potential, in atomic units, is given by:
\be
\label{Potential}
V_H(r) = - {\d e^{-\d r}\o 1-e^{-\d r}},
\ee
where $\d$ is the screening parameter. This potential has been used in several
branches of Physics, (see \cite{Varshni} and references therein).

The Schroedinger equation for this potential can be solved in closed form for
$s$ waves, \cite{Lan}.  However, for $l\not=0$ it is necessary to use
numerical methods, \cite{Varshni}.

Supersymmetric Quantum Mechanics (SQM) has been used to solve Schroedinger
equation of solvable potentials, \cite{Levai}, partially solvable ones,
\cite{Drigo1}, in the WKB-approximation, \cite{Dutt}, and it has also been
applied in variational method, as recently suggested, \cite{Cooper1},
\cite{Drigo2}.

The supersymmetric formalism has already been used to study some aspects
of the Hulth\'en potential, \cite{Laha}, \cite{Talukdar}. Here, the exact
analytical solution for this potential is reobtained, for $l=0$ to show the
consistency of the method. When  $l \neq 0$ we interpret that the supersymmetry
is ``broken" by the potential barrier terms.  Nonetheless, the supersymmetry
gives us eigenfunctions that allow us to compute the eigenvalues of the
variational method.  The eigenfunctions for $2p$ and $3d$ states are evaluated
for some values of the parameter delta.  Our results are compared with direct
numerical integration data, \cite{Varshni}.

\section{The Hulth\'en Potential with $l=0$}

The Hamiltonian for the Hulth\'en potential ($l=0$) can be written as:
\be
\label{Hamiltonian}
H_1 = \hm{d^2 \o dr^2} - {\d e^{-\d r}\o 1-e^{-\d r}}.
\ee
{}From the Hamiltonian hierarchy we can obtain its eigenfunctions and
eigenvalues, \cite{Sukumar}, \cite{Drigo3}.  In this way, first we factorise
the
Hamiltonian (\ref{Hamiltonian}):
\be
\label{H_1}
H_1 - E_0^{(1)} = a_1^+ a_1^-
\ee
with creation and anihilation operators defined by
\be
\label{aoperator}
a_1^{\pm} = {1\over \sqrt 2}({\mp}{d\o dr} + W_1(r))
\ee
and we determine the ground state eigenvalue $E_0^{(1)}$, the superpotential $
W_1(r)$ and the ground state eigenfunction
\be
\Psi_0^{(1)} \propto e^{-\int_0^r  W_1(r') dr'}.
\ee
{}From the operators $a_1^+$ and $a_1^-$, we write $H_2$,the supersymmetric
partner of $H_1$ as
\be
H_2 - E_0^{(1)} = a_1^- a_1^+
\ee
and factorise $H_2$ in an analogous way to (\ref{H_1}),
\be
\label{a2operator}
a_2^+ a_2^- = H_2 - E_0^{(2)}
\ee
with $a_2^{\pm}$ defined similarly to (\ref{aoperator}). We then determine
$E_0^{(2)}$, $W_2(r)$ and $\Psi_0^{(2)}$.  Interchanging the operators $a_2^+$
and $a_2^-$ in (\ref{a2operator}) we obtain the supersymmetric partner of
$H_2$.  Repeating this process $n$ times we obtain the eigenvalues and
eigenfunctions for the $n$-th ground state Hamiltonian. The superalgebra allows
us to relate these results with the original Hamiltonian by the relations:
\be
\label{suzy}
E_n^{(1)} = E_0^{(n+1)},\;\;\;\;\;\Psi_n^{(1)}(r) =
 a_1^+a_2^+...a_n^+\Psi_0^{(n+1)}(r).
\ee
In our case, using the Hulth\'en potential, the $n$-th superpotential is given
by \be
W_n(r) = - {n\d e^{-\d r}\o 1-e^{-\d r}} + {1\o n} - {n\o 2}\d
\ee
that corresponds to the $n$-th member of the Hamiltonian hierarchy:
\be
\label{potential}
V_n(r) = W_n^2(r) -{d\o dr}W_n(r) =
{n(n-1){\d}^2 e^{-2\d r}\o 2(1-e^{-\d r})^2} -{[n(1-n)\d +2]\d e^{-\d r}\o
2(1-e^{-\d r})}.
\ee
The energy-eigenvalue and eigenfunction, eq.(\ref{suzy}), are
\be
\label{eigenvalues}
E_n^{(1)} = {1\o 2} (-{n\over 2} \d +{1\over n})^2 \\
\ee
\be
\Psi_0^{(n)}(r) = (1 - e^{-\d r})^n e^{-[ {1\o n} - { n\o 2}\d]r}.
\ee
We can verify that the energy-eigenvalues (\ref{eigenvalues}) are the same
given
in reference \cite{Lan}.
\section {The Hulth\'en Potential with $l\not=0$ }

The Hamiltonian for the Hulth\'en potential when $l\not= 0$ is written as:
\be
\label{Hulthen}
H = \hm{d^2 \o dr^2} - {\d e^{-\d r}\o 1-e^{-\d r}} + {l(l+1) \o 2 r^2}.
\ee

The potential barrier term prevent us to build the superfamily as in the
$l = 0$ case, since the potential is not exactly solvable.  However, several
numerical approaches have been used in order to evaluate the spectra of
energy-eigenvalues and eigenfunctions.  In particular, Greene and Aldrich,
\cite{Greene}, suggested an Hulth\'en effective potential in which the
eigenfunctions are used as variational trial wave functions.

Based in our results for the case $l = 0$, we introduce a new effective
potential whose functional form is suggested by eq.(\ref{potential}),
\be
\label{effective}
V_{eff}(r) = - {\d e^{-\d r}\o 1-e^{-\d r}} + {l(l+1)\o 2}{{\d}^2 e^{-2\d r}\o
(1-e^{-\d r})^2}.
\ee

We note that, for small values of  $\d$, the second term of (\ref{effective})
 gives us a potential barrier term of (\ref{Hulthen}) in first approximation.
This potential differs from the one used in \cite{Greene} by the exponential
numerator which is a quadratic term. The advantage of using the effective
potential (\ref{effective}) is the fact that we can vary its parameters
without changing its functional form.  This is reinforced by the constructive
method of determining wave functions based on supersymmetry.

As the effective potential given by (\ref{effective}) has the same functional
form  as (\ref{potential}), we can solve the Schroedinger problem by the
factorisation method of SQM and find the whole super family. The
superpotential of the $n$-th member of the family is given by
\be
W_n(r) = B_n { e^{-\d r}\o 1 - e^{-\d r}} + C_n
\ee
where
\be
B_n = - {1 \o 2} (\d + \sqrt {{\d}^2 + 4B_{n-1} (B_{n-1} - \d)}, \;\;\;\;\;
C_n = - {B_{n-1} (\d - 2 C_{n-1}) - B_n \d \o 2 B_n},
\ee
and the energy and wave functions are given by (\ref{suzy}), where
\be
B_1 = - {\d \o 2} (1 + \sqrt {1 + 4l(l+1)}), \;\;\;\;\; C_1 = - {\d \o 2}
{B_1 + 2\o B_1},\;\;\;\; B_0 = 0.
\ee
The energy eigenvalues and wave functions given by (\ref{suzy}) are
\be
E_0^{(n+1)} = {1 \o 2} C_{n+1}^2
\ee
\be
\label{Psidelta}
\Psi_0^{(1)} = (1 - e^{-\d r})^{-{B_1 \o \d}} e^{-C_1 r}.
\ee
Notice that $B$'s and $C$'s depend on $l$.

Thus, the spectrum of the effective potential (\ref{effective}) has quantum
number $n = 0, 1, ...$ that shows us which member of the superfamily is going
to be used. For instance, $n = 0$ and $l = 1$ corresponds to state $2p$;  $n =
1$ and $l = 1$ corresponds to state $3d$ and so on.  Fixing $l = 0$ we recover
the original Hulth\'en potential, as expected.

We now look at the Hulth\'en potential (\ref{Hulthen})  and
solve the Schroedinger problem by the variational method.  We start from
$V_{eff}$ given by  (\ref{effective}) whose eigenfunctions are given by
(\ref{suzy}) and (\ref{Psidelta}). For the state $2p$ we use the first member
of the superfamily ($l=1$) as our trial wave function, changing $\d$ by the
variational parameter $\mu$, i.e.,
\be
\Psi_{\mu} = \Psi_0^{(1)}(r,\mu) = (1 - e^{-\mu r})^{-{B_1 \o \mu}} e^{-C_1 r}.
\ee
The energy  is obtained by minimisation with respect to $\mu$. Thus, the
equation to be $\;\;$ minimised is
\be
\label{energymu}
E_{\mu} = {\int_0^{\infty} \Psi_{\mu}(r) [\hm {d^2 \o dr^2} - {\d e^{-\d r}\o
1-e^{-\d r}} + {l(l+1)\o 2r^2}] \Psi_{\mu}(r) dr
\o \int_0^{\infty} \Psi_{\mu}(r)^2 dr}.
\ee

The second derivative in (\ref{energymu}) can be evaluated analitically.
However, the integration has to be carried out numerically. Our explicit values
for the $2p,\; (l=1)$ and $3d,\; (l=2)$ energy states for some values of the
parameter $\d$ are listed bellow  in Table 1. They are shown together with
direct numerical integration data.
\vskip 1cm
\eject
Table 1. Energy eigenvalues as a function of the screening parameter for the
states $2p$ and $3d$, [eq.(\ref{energymu})]. Comparison is made with numerical
data of
Ref.\cite{Varshni}.
\vskip 1cm

\begin{tabular}{|l|c|c|c|c|c|} \hline
\multicolumn{1}{|l}{State} &
\multicolumn{1}{|c} {Delta} &
\multicolumn{1}{|c} {Variational result}   &
\multicolumn{1}{|c|} {Numerical Integration }  \\  \hline
2p & 0.025 & -0.112760  & -0.1127605   \\ \hline
 & 0.050 & -0.101042  & -0.1010425   \\ \hline
 & 0.075 & -0.089845  & -0.898478  \\ \hline
 & 0.100 & -0.079170  & -0.0791794  \\ \hline
 & 0.150 & -0.059495   & -0.0594415 \\ \hline
 & 0.200 & -0.041792   & -0.0418860 \\ \hline
3p & 0.025 & -0.043601  & -0.0437069   \\ \hline
 & 0.050 & -0.032748  & -0.0331645   \\ \hline
 & 0.075 & -0.023010  & -0.0239397  \\ \hline
 & 0.100 & -0.014433  & -0.0160537  \\ \hline
\end{tabular}

\eject
\section {Conclusions}

We have obtained from the formalism of SQM the exact analytical eigenfunction
and energy eigenvalue for the Hulth\'en potential for $l = 0$.  When $l \neq
0$,
we used an effective potential suggested by the case $l = 0$,
to give us a variational trial wave function. The energies for the $2p$ and
$3d$ states were obtained for some values of the parameter $\d$.

We have used the same variational trial wave function for the $2p$ and $3d$
states, eq.(\ref{Psidelta}). The only change was the angular moment number
$l$.  However, as the suggested potential has the same functional form, we
could determine all the eigenfunctions and therefore we could  use the most
appropriate in each case, (see \cite{Chuan}).

We note that better results have been obtained for the $2p$ state, for small
values of $\d$. This is expected since for small values of $\d$ the
effective potential (\ref{effective}) becomes closer to the original Hulth\'en
potential (\ref{Hulthen}), and for $l$ small the contribution of this angular
moment term in the potential is also small. The advantage of this process is
that only the form of the potential (\ref{effective}) is important to obtain
the eigenfunctions; the multiple parameters can be changed conveniently in
order to obtain better results.

The algorithm used here is very powerful because it gives us a systematic
method to look for variational trial wave functions based on the functional
form of the effective potential in the factorisation method of SQM.

\end{document}